\documentclass[conference]{IEEEtran}
\IEEEoverridecommandlockouts
\usepackage{cite}
\usepackage{amsmath,amssymb,amsfonts}
\usepackage{algorithmic}
\usepackage[pdftex]{graphicx} 
\usepackage{textcomp}
\usepackage{xcolor}
\usepackage{bm}
 \usepackage{algorithm}
\usepackage{algorithmic}
\def\BibTeX{{\rm B\kern-.05em{\sc i\kern-.025em b}\kern-.08em
    T\kern-.1667em\lower.7ex\hbox{E}\kern-.125emX}}
\begin{document}

\title{Deep Reinforcement Learning Based \\ 
Networked Control with Network Delays \\
for Signal Temporal Logic Specifications
}

\author{\IEEEauthorblockN{Junya Ikemoto}
\IEEEauthorblockA{\textit{Engineering Science} \\
\textit{Osaka University}\\
Toyonaka, Japan \\
email address: ikemoto@hopf.sys.es.osaka-u.ac.jp}
\and
\IEEEauthorblockN{Toshimitsu Ushio}
\IEEEauthorblockA{\textit{Engineering Science} \\
\textit{Osaka University}\\
Toyonaka, Japan \\
email address: ushio@sys.es.osaka-u.ac.jp}
}

\maketitle

\begin{abstract}
We apply deep reinforcement learning (DRL) to design of a networked controller with network delays to complete a temporal control task that is described by a signal temporal logic (STL) formula. STL is useful to deal with a specification with a bounded time interval for a dynamical system. In general, an agent needs not only the current system state but also the past behavior of the system to determine a desired control action for satisfying the given STL formula. Additionally, we need to consider the effect of network delays for data transmissions. Thus, we propose an extended Markov decision process (MDP) using past system states and control actions, which is called a $\tau d$-MDP, so that the agent can evaluate the satisfaction of the STL formula considering the network delays. Thereafter, we apply a DRL algorithm to design a networked controller using the $\tau d$-MDP. Through simulations, we also demonstrate the learning performance of the proposed algorithm.
\end{abstract}


\section{INTRODUCTION}
\begin{figure}
  \centering
  \includegraphics[width=9.0cm]{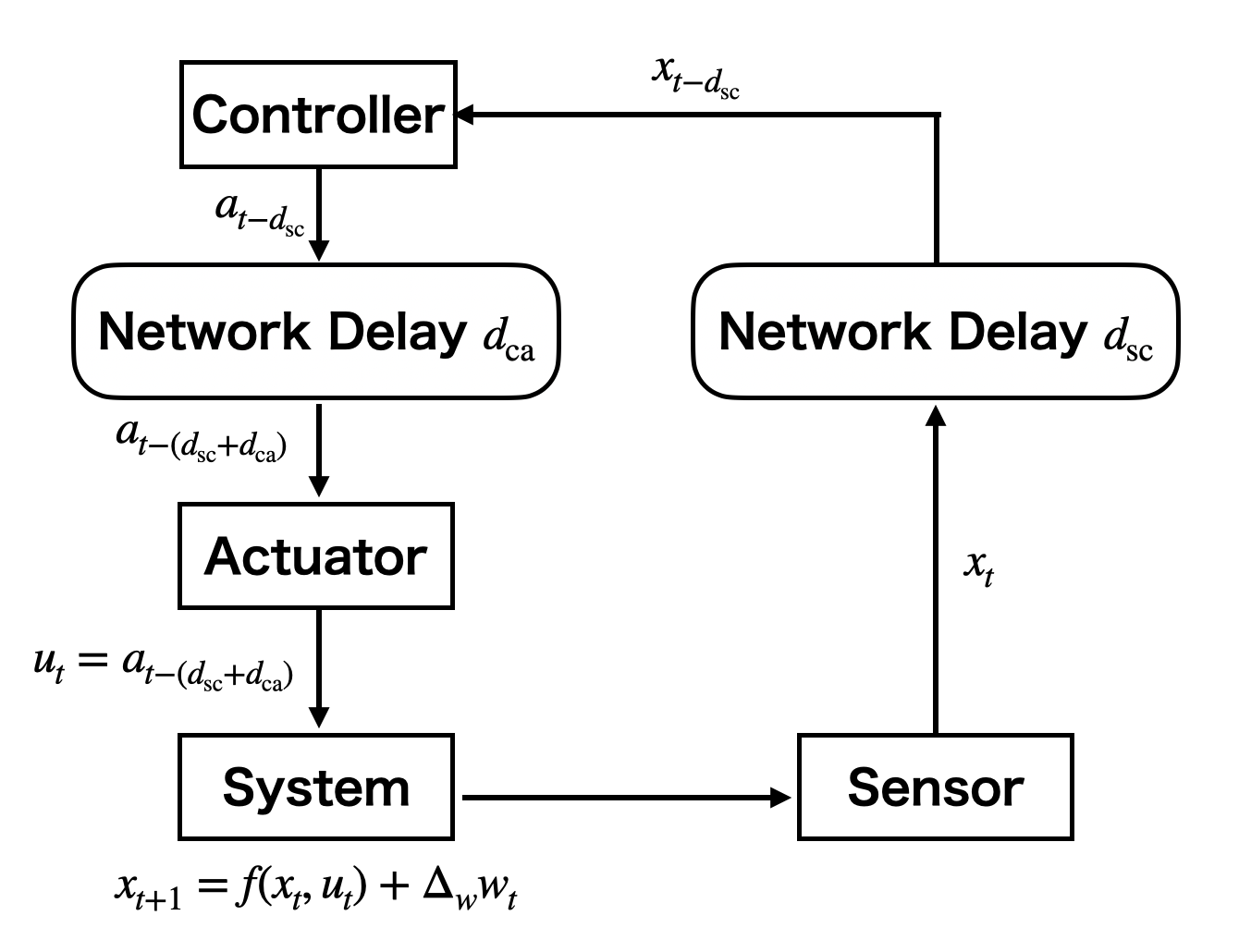}
  \caption{Illustration of a network control system (NCS). Although the system has many advantages, there exist two types of network delays that may degrade control performance.}
  \label{NCSs}
\end{figure}

\it Networked control systems \rm (NCSs) have attracted considerable attention owing to the development of network technologies \cite{NCS}. NCSs are systems with loops closed through networks, as shown in Fig.\ \ref{NCSs}, and have many advantages in various control problems. However, in NCSs, network delays are caused by data transmission between a sensor/actuator and a controller. In conventional model-based controller designs, we identify the mathematical model of a system and network delays beforehand. However, it may be difficult to identify them precisely in real-world problems. Subsequently, \it reinforcement learning \rm (RL) \cite{RL_Sutton} is useful because we can adaptively design a controller through interactions with the system.

RL is a machine learning method used in various fields to solve sequential decision-making problems, and has been studied in the control field because it is strongly associated with optimal control methods from a theoretical point of view. Moreover, RL with \it deep neural networks \rm (DNNs), called \it Deep RL \rm (DRL), has been developed for complicated decision-making problems \cite{DRL_Book}, such as playing Atari 2600 video games \cite{DQN} and locomotion or manipulation of complicated systems \cite{TRPO,DDPG,TD3,SAC}. DRL-based networked controller designs have been proposed \cite{Event_trigger,Deep_CAS,DRL_NCSs_delay}. In \cite{Event_trigger}, Baumann \it et al.\ \rm proposed a DRL-based event-triggered control method. In \cite{Deep_CAS}, Demirel \it et al.\ \rm proposed a control-aware scheduling algorithm to synthesize an optimal controller for some subsystems. In \cite{DRL_NCSs_delay}, we proposed DRL-based networked controller designs to stabilize an uncertain nonlinear system with network delays. 

On the other hand, in RL-based controller designs, we must design a reward function for desired system behavior beforehand, which is difficult for a temporal high-level control task. To handle the temporal control task, \it temporal logic \rm (TL) \cite{model_checking} is useful. TL is a branch of formal methods and has also been applied to several control problems \cite{Belta_FM_discrete_time_systems}. \it Signal temporal logic \rm (STL) \cite{STL} is particularly useful in designing controllers for dynamical systems as it can specify continuous signals within a bounded time interval. STL has also been studied in the machine learning community. In \cite{STL_net}, Ma \it et al\rm. proposed an STL-based learning framework with knowledge of model properties. Moreover, RL-based controller design methods using STL formulae have been proposed \cite{STL_Q_learning,Practical_STL_Q_learning,STL_reward_shaping,MBRL_STL}. In \cite{STL_Q_learning}, Aksaray \it et al.\ \rm proposed a Q-learning-based method to design a control policy that satisfies a given STL specification. They introduced an extended \it Markov decision process \rm (MDP), which is called a $\tau$\it -MDP\rm, and designed a reward function to learn a control policy satisfying the STL specification. The extended state of the $\tau$-MDP comprises the current state and the past system state sequence, where the dimension of the extended state depends on the given STL formula. In \cite{Practical_STL_Q_learning}, Venkataraman \it et al.\ \rm proposed a tractable learning method using a flag state instead of the past state sequence to mitigate the \it curse of dimensionality\rm. However, these methods cannot be directly applied to problems with continuous state and action spaces because they are based on tabular Q-learning. In \cite{STL_reward_shaping}, Balakrishnan \it et al.\ \rm introduced a \it partial signal \rm and proposed a DRL-based method. In \cite{MBRL_STL}, Kapoor \it et al.\ \rm proposed a model-based DRL method. The model of the system was learned using a DNN, and the controller was designed using a nonlinear model predictive control method. Li \it et al.\ \rm proposed a policy search algorithm using \it truncated linear temporal logic \rm (TLTL) that does not have a time bound \cite{TLTL_1,TLTL_2}. 

In this study, we formulate a temporal control specification as an STL formula and propose a DRL-based networked controller design in the presence of networked delays. 

\textbf{Contribution:} The main contribution of this paper is the development of a DRL-based networked controller design for satisfying STL specifications with fixed network delays. In this study, it is assumed that we cannot identify the mathematical model of the system and the network delays beforehand, where the bounds of the network delays are known. To design the networked controller, we proposed an extended MDP, which is called a $\tau d$-\it MDP\rm, and a practical DRL-based networked controller design using the extended MDP. 

\textbf{Structure:} The remainder of this paper is organized as follows. In Section II, we review STL as preliminaries. In Section III, we formulate a networked control problem for a stochastic discrete-time system. In Section IV, we propose a DRL-based networked controller design to satisfy a given STL specification in the presence of networked delays. In Section V, using numerical simulations, we demonstrate the usefulness of the proposed method. In Section VI, we conclude the paper and discuss future work. 

\textbf{Notation:} $\mathbb{N}_{\ge0}$ is the set of non-negative integers. $\mathbb{R}$ is the set of the real numbers. $\mathbb{R}_{\ge0}$ is the set of non-negative real numbers. $\mathbb{R}^{n}$ is the $n$-dimensional Euclidean space. $0_{n}$ is an $n$-dimensional zero vector. For a set $A\subseteq\mathbb{R}$, $\max A$ and $\min A$ are the maximum value and the minimum value in $A$ if they exist, respectively.

\section{Signal Temporal Logic}
In this study, we describe a desired control task as an STL formula with the following \textit{syntax}. 
\begin{eqnarray}
\Phi&::=&G_{[0,T_e]}\phi\ |\ F_{[0,T_e]}\phi,\nonumber\\
\phi&::=&G_{[t_s,t_e]}\varphi\ |\ F_{[t_s,t_e]}\varphi\ |\ \phi\land\phi\ |\ \phi\lor\phi,\nonumber\\
\varphi&::=&\psi\ |\ \lnot\varphi\ |\ \varphi\land\varphi\ |\ \varphi\lor\varphi\nonumber,
\end{eqnarray}
where $T_e$, $t_s$, and $t_e\in\mathbb{N}_{\ge0}$ are nonnegative constants for the time bounds of temporal operators. $\Phi$, $\phi$, $\varphi$, and $\psi$ are STL formulae. $\psi$ is a predicate in the form of $h(x)\le y$, where $h:\mathcal{X}\to\mathbb{R}$ is a function of a system state, and $y\in\mathbb{R}$ is a constant. The Boolean operators $\lnot$, $\land$, and $\lor$ are \textit{negation}, \textit{conjunction}, and \textit{disjunction}, respectively. The temporal operators $G_{\mathcal{T}}$ and $F_{\mathcal{T}}$ refer to \textit{Globally} (always) and \textit{Finally} (eventually), respectively, where $\mathcal{T}$ denotes the time bound of the temporal operator in the form of $[t_{s},t_{e}],\ t_{s}\le t_{e}$. $\phi_{i}=G_{[t_s^i,t_e^i]}\varphi_{i}$ or $F_{[t_s^i,t_e^i]}\varphi_{i},\ i\in\{1,2,...,M\}$ are called \textit{STL sub-formulae}. $\phi$ comprises multiple STL sub-formulae $\{\phi_i\}_{i=1}^{M}$.

$x_{t}$ and $x_{t_1:t_2}$ denote the state at $t$ and the partial trajectory for a discrete-time interval $[t_1,t_2]$, where $t_1\le t_2$. The \textit{Boolean semantics} of STL is recursively defined as follows:
\begin{eqnarray}
&& x_{t:T}\models\psi\Leftrightarrow h(x_{t})\le y,\nonumber\\
&& x_{t:T}\models\lnot\psi\Leftrightarrow \lnot(x_{t:T}\models\psi),\nonumber\\
&& x_{t:T}\models\phi_{1}\land\phi_{2}\Leftrightarrow x_{t:T}\models\phi_{1}\land x_{t:T}\models\phi_{2},\nonumber\\
&& x_{t:T}\models\phi_{1}\lor\phi_{2}\Leftrightarrow x_{t:T}\models\phi_{1}\lor x_{t:T}\models\phi_{2},\nonumber\\
&&x_{t:T}\models G_{[t_s,t_e]}\phi\Leftrightarrow x_{t':T}\models\phi,\ \forall t'\in [t+t_{s},t+t_{e}]\nonumber\\
&&x_{t:T}\models F_{[t_s,t_e]}\phi\Leftrightarrow \exists t'\in [t+t_s,t+t_e],\ \text{s.t. } x_{t':T}\models\phi,\nonumber
\end{eqnarray}
where $T\ (\ge t_e)$ denotes the length of the trajectory.

The \textit{quantitative semantics} of STL is recursively defined as follows:
\begin{eqnarray}
\rho(x_{t:T},\psi)&=&y-h(x_{t}),\nonumber\\
\rho(x_{t:T},\lnot\psi)&=&-\rho(x_{t:T},\psi)\nonumber\\
\rho(x_{t:T},\phi_{1}\land\phi_{2})&=&\min\{\rho(x_{t:T},\phi_{1}),\rho(x_{t:T},\phi_{2})\},\nonumber\\
\rho(x_{t:T},\phi_{1}\lor\phi_{2})&=&\max\{\rho(x_{t:T},\phi_{1}),\rho(x_{t:T},\phi_{2})\},\nonumber\\
\rho(x_{t:T},G_{[t_s,t_e]}\phi)&=&\min_{t'\in [t+t_s,t+t_e]}\rho(x_{t':T},\phi),\nonumber\\
\rho(x_{t:T},F_{[t_s,t_e]}\phi)&=&\max_{t'\in [t+t_s,t+t_e]}\rho(x_{t':T},\phi),\nonumber
\end{eqnarray}
which quantifies how well the trajectory satisfies a given STL formulae \cite{Robust_degree_STL}.

The \textit{horizon length} of an STL formula is recursively defined as follows:
\begin{eqnarray}
\text{hrz}(\psi)&=&0,\nonumber\\
\text{hrz}(\phi)&=& t_{e},\ \text{for }\phi=G_{[t_{s},t_{e}]}\varphi,\ \text{or }F_{[t_s,t_e]},\varphi\nonumber\\
\text{hrz}(\lnot\phi)&=&\text{hrz}(\phi),\nonumber\\
\text{hrz}(\phi_1\land\phi_2)&=&\max\{\text{hrz}(\phi_1),\text{hrz}(\phi_2)\},\nonumber\\
\text{hrz}(\phi_1\lor\phi_2)&=&\max\{\text{hrz}(\phi_1),\text{hrz}(\phi_2)\},\nonumber\\
\text{hrz}(G_{[t_s,t_e]}\phi)&=&t_e+\text{hrz}(\phi),\nonumber\\
\text{hrz}(F_{[t_s,t_e]}\phi)&=&t_e+\text{hrz}(\phi).\nonumber
\end{eqnarray}
$\text{hrz}(\phi)$ is the required length of the state sequence to verify the satisfaction of the STL formula $\phi$ \cite{Horizon_length}.

\section{Problem Statement}
We design a networked controller for the following stochastic discrete-time dynamical system as shown in Fig.\ \ref{NCSs}.
\begin{eqnarray}
x_{t+1}=f(x_{t},u_{t})+\Delta_{w}w_{t},\label{dynamical_system}
\end{eqnarray}
where $x_{t}\in\mathcal{X}$, $u_{t}\in\mathcal{U}$, and $w_{t}\in\mathcal{W}$ are the system state, the control input, and the system noise at the discrete-time $t\in\{0,1,...\}$. $\mathcal{X}=\mathbb{R}^{n_x}$, $\mathcal{U}\subseteq\mathbb{R}^{n_u}$, and $\mathcal{W}=\mathbb{R}^{n_x}$ are the state space, the control input space, and the system noise space, respectively. The system noise $w_{t}$ is an independent and identically distributed random variable with a probability density $p_{w}:\mathcal{W}\to\mathbb{R}_{\ge0}$. $\Delta_{w}$ is a regular matrix that is a weighting factor of the system noise. $f:\mathcal{X}\times\mathcal{U}\to\mathcal{X}$ is a function that describes the system dynamics. Then, we have the transition probability density $p_{f}(x'|x,u):=|\Delta_{w}^{-1}|p_{w}(\Delta_{w}^{-1}(x'-f(x,u)))$. The initial state $x_0\in\mathcal{X}$ is sampled from a probability density $p_{0}:\mathcal{X}\to\mathbb{R}_{\ge0}$. The goal is to design a control policy that satisfies $x_{0:T}^{\pi}\models\Phi $, where $x_{0:T}^{\pi}$ is a system trajectory controlled by a control policy $\pi$ and $\Phi$ is a given STL specification.

In the NCS, there exist two types of network delays: a \textit{sensor-to-controller delay} $d_{\text{sc}}\in\mathbb{N}$ caused by the transmission of the observed state and a \textit{controller-to-actuator delay} $d_{\text{ca}}\in\mathbb{N}$ caused by the transmission of a control input computed by the controller. In this study, it is assumed that these delays are uncertain constants bounded by the maximum delays $d_{\text{sc}}^{\max}\in\mathbb{N}$ and $d_{\text{ca}}^{\max}\in\mathbb{N}$, respectively. Then, the controller computes the $k$-th control input $a_{k}$ based on the $k$-th observed state $x_{k}$ at $t=k+d_{\text{sc}}$. Actually, the control input $a_{k}$ is input to the system as follows:
\begin{eqnarray}
u_{t}=\begin{cases}
	a_{k} & t = k + d_{\text{sc}} + d_{\text{ca}},\\
	0_{n_u} & t< d_{\text{sc}} + d_{\text{ca}},\label{delayed_input}
\end{cases}
\end{eqnarray}
where $0_{n_u}$ is a zero-vector of $\mathbb{R}^{n_u}$, that is the actuator inputs the control input $0_{n_u}$ until receiving $a_{0}$. Note that the controller cannot control the system until $t=d_{\text{sc}}+d_{\text{ca}}$. The controller computes control inputs $a_{0},a_{1},...,a_{T-d_{\text{sc}}-d_{\text{ca}}}$ to satisfy the STL specification.

Furthermore, it is assumed that the mathematical models $f$ and $p_{w}$ are unknown. Thus, we apply RL to design a networked controller for satisfying the given STL specification $\Phi$. In RL, an agent interacts with an environment and learns its control policy using the past interaction data. In this study, we regard the controller and everything outside the controller as the agent and the environment for RL, respectively. A control input determined by the agent is called a \textit{control action}. However, a standard RL algorithm cannot be directly applied due to the following issues.
\begin{enumerate}
\renewcommand{\labelenumi}{(\roman{enumi})}
\item The desired control action at each step in order to satisfy the STL specification $\Phi$ is determined not only by the current state but also by the past system behavior.
\item We must design a reward function to evaluate the satisfaction of the STL specification appropriately.
\item The classical RL algorithm cannot deal with continuous state and action spaces directly.
\item There exist uncertain network delays in the NCS. 
\end{enumerate}
In the next section, we propose a DRL-based controller design that resolves the issues.

\section{DRL-BASED NETWORKED CONTROLLER DESIGN FOR STL SPECIFICATIONS}

\begin{figure*}[h]
\begin{center}
  \includegraphics[width=16.0cm]{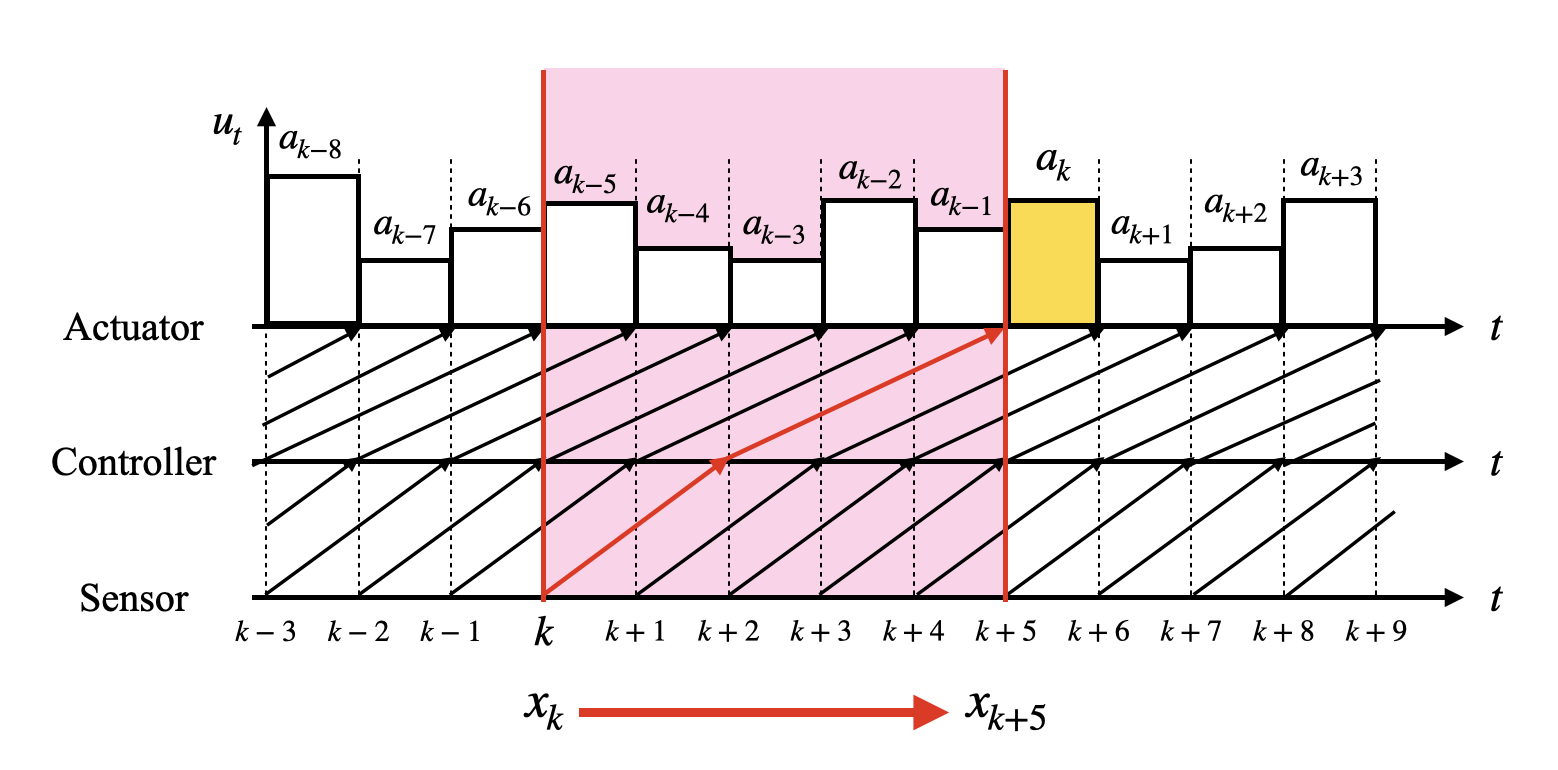}
  \caption{Illustration of the network delays in data transmissions for the worst case, where $d_{\text{sc}}^{\max}=2$ and $d_{\text{ca}}^{\max}=3$ ($d=5$). At $t=k+2$, the agent predicts $x_{k+5}$ and determines $a_{k}$ using the $k$-th state $x_{k}$ and the past actions $a_{k-5},...,a_{k-1}$.}
  \label{NCS_time_table}
\end{center}
\end{figure*}

\subsection{$\tau d$-Markov decision process}
For issue (i), Aksaray \it et al.\ \rm introduced an extended MDP, which is called a $\tau$-MDP, using a finite state sequence in \cite{STL_Q_learning}. For issue (ii), they designed a reward function of the $\tau$-MDP to satisfy a given STL specification using the \textit{log-sum-exp} approximation. They apply the classical Q-learning to design a policy for satisfying the given STL formula. However, their method cannot handle continuous control tasks directly. To resolve issue (iii), we extended the method using a DRL algorithm for problems with a continuous state-action space. In this study, we apply the \textit{soft actor critic} (SAC) algorithm, because it has better sample efficiency and asymptotic performance. Additionally, we must consider network delays. For issue (iv), in \cite{DRL_NCSs_delay}, we proposed an extended state that comprises a current system state and previously determined control actions. As shown in Fig.\ \ref{NCS_time_table}, we consider the worst case scenario. At $t=k$, the sensor observes the $k$-th system state $x_{k}$, which is transmitted to the agent (controller) through the network. The agent receives the observed state $x_{k}$ and determines the $k$-th control action $a_{k}$ at $t=k+d_{\text{sc}}^{\max}$. The action $a_{k}$ is sent to the actuator through the network. The actuator receives the $k$-th control action $a_{k}$ and updates the control input $u_{t}=a_{k}$ at $t=k+d\ (:=k+d_{\text{sc}}^{\max}+d_{\text{ca}}^{\max})$. Then, it is desirable for the agent to predict the future state $x_{k+d}$ with available information and determine the $k$-th control action $a_{k}$ based on the prediction. If we are aware of the system dynamics (\ref{dynamical_system}), we can predict the future state. However, the prediction requires not only the system dynamics (\ref{dynamical_system}) but also the information $u_{t},\ t\in[k,k+d]$, which is the past control action sequence $a_{k-d:k-1}=a_{k-d},\ a_{k-d+1},\ ...,\ a_{k-1}$. Thus, we use not only the extended state proposed in \cite{STL_Q_learning} but also previously determined control actions. Actually, the true network delays $d_{\text{sc}}$ and $d_{\text{ca}}$ are uncertain constants. The agent adapts the true network delays through interactions with the system using sufficient information for predictions in the worst case. For issues (i), (ii), (iii), and (iv), we remodel the interactions between the agent and the system as the following extended MDP, which is called a $\tau d$-\textit{MDP}.

\noindent \textbf{Definition 1 ($\tau d$-MDP):} Given an STL formula $\Phi=G_{[0,T_{e}]}\phi\ \text{or}\ F_{[0,T_{e}]}\phi$, where $\text{hrz}(\Phi)=T$ and $\phi$ comprises multiple STL sub-formulae $\phi_{i},\ i\in\{1,2,...,M\}$. Subsequently, we set $\tau=\text{hrz}(\phi)+1$, that is, $T=T_{e}+\tau-1$. It is assumed that $d_{\text{sc}}^{\max}+d_{\text{ca}}^{\max}=d$. A $\tau d$-MDP is defined by a tuple $\mathcal{M}_{\tau,d}=\left<\mathcal{Z},\mathcal{U},p_{0}^{z},p^{z},R^{z} \right>$, where
\begin{itemize}
\item $\mathcal{Z}\subseteq\mathcal{X}^{\tau}\times\mathcal{U}^{d}$ is an extended state space. Each extended state is denoted by $z=[(x^{\tau})^{\top}\ (a^{d})^{\top}]^{\top}$, where $x^{\tau}=[x^{\tau}[0]^{\top}\ x^{\tau}[1]^{\top}\ ...\ x^{\tau}[\tau-1]^{\top}]^{\top}$ and $a^{d}=[a^{d}[0]^{\top}\ a^{d}[1]^{\top}\ ...\ a^{d}[d-1]^{\top}]^{\top}$ are a past state sequence and a previously determined control action sequence, respectively, that is, $x^{\tau}[i]\in\mathcal{X},\ \forall i\in\{0,1,...,\tau-1\}$ and $a^{d}[j]\in\mathcal{U},\ \forall j\in\{0,1,...,d-1\}$.
\item $\mathcal{U}$ is a control action space.
\item $p_{0}^{z}$ is a probability density for an initial extended state $z_{0}=[(x_{0}^{\tau})^{\top}\ (a_{0}^{d})^{\top}]^{\top}$ with $x_{0}^{\tau}[i]=x_{0},\ \forall i\in\{0,1,...,\tau-1\}$ and $a_{0}^{d}[j]=0_{n_{u}},\ \forall j\in\{0,1,...,d-1\}$, where $x_{0}$ is generated from $p_{0}$.
\item $p^{z}$ is a transition probability density for the extended state $z$. In the case where the system state is updated by $x'\sim p_{f}(\cdot|x,u)$, the extended state is updated by $z'\sim p^{z}(\cdot|z,a)$ as follows: 
\begin{eqnarray}
&&a^{d'}[j]=a^{d}[j+1],\ \forall j\in\{0,1,...,d-2\},\nonumber\\
&&a^{d'}[d-1]=a, \nonumber\\
&&x^{\tau'}[i]=x^{\tau}[i+1],\ \forall i\in\{0,1,...,\tau-2\},\nonumber\\ 
&&x^{\tau'}[\tau-1]\sim p_{f}(\cdot|x^{\tau}[\tau-1],a^{d'}[d-1-d_{\text{sc}}-d_{\text{ca}}]),\nonumber
\end{eqnarray}
where $z=[(x^{\tau})^{\top}\ (a^{d})^{\top}]^{\top}$ and $z'=[(x^{\tau'})^{\top}\ (a^{d'})^{\top}]^{\top}$ are the current extended state and the next extended state, respectively.
\item $R_{z}:\mathcal{Z}\to\mathbb{R}$ is a reward function. Based on \cite{STL_Q_learning}, it is defined by
\begin{eqnarray}
&&R_{z}(z)\nonumber\\
&&=\begin{cases}
	-\exp(-\beta \bm{1}(\rho(x^{\tau},\phi))) & \text{for } \Phi=G_{[0,T_{\text{e}}]}\phi, \\
	\exp(\beta \bm{1}(\rho(x^{\tau},\phi))) & \text{for } \Phi=F_{[0,T_{\text{e}}]}\phi,	
\end{cases}\nonumber\\
\label{reward_function}
\end{eqnarray}
where $\beta>0$ is a reward parameter. The function $\bm{1}:\mathbb{R}\to\{0,1\}$ is an indicator defined by
\begin{eqnarray}
\bm{1}(y) = \left\{ 
	\begin{array}{ll}
		1 & \mbox{if } y \geq 0 , \\ 
		0 & \mbox{if } y < 0 .
	\end{array}\right. \nonumber
\end{eqnarray}
The reward function is designed for satisfying the given STL specification using the log-sum-exp approximation \cite{STL_Q_learning}. 
\end{itemize}

The agent determines a control action according to a stochastic policy $\pi:\mathcal{Z}\to\mathcal{P}(\mathcal{U})$, where $\mathcal{P}(\mathcal{U})$ denotes the set of probability distributions over $\mathcal{U}$. In the SAC algorithm \cite{SAC}, we use the objective with the entropy term as follows:
\begin{eqnarray} 
J(\pi)&=&E_{p_{\pi}}\left[\sum_{k=0}^{T-d_{\text{sc}}-d_{\text{ca}}}\gamma^{k}(R_{z}(z_{k})+\alpha\mathcal{H}(\pi(\cdot|z_{k})))\right],\nonumber
\end{eqnarray} 
where $\gamma\in[0,1)$ is a discount factor, $p_{\pi}$ is a trajectory distribution by the policy $\pi$, $\mathcal{H}$ is the entropy defined by $\mathcal{H}(\pi(\cdot|z))=E_{a\sim \pi(\cdot|z)}\left[-\log \pi(a|z)\right]$, and $\alpha>0$ is an entropy temperature. The goal is to obtain a control policy $\pi$ that maximizes the objective. We give the stochastic policy $\pi$ using a Gaussian with the mean $\mu_{\theta_{\pi}}$ and the standard deviation $\sigma_{\theta_{\pi}}$ output by a DNN with \textit{reparameterization trick} \cite{VAE}, which is called an \textit{actor DNN}, whose parameter vector is denoted by $\theta_{\pi}$. Additionally, we need to estimate the objective $J(\pi)$. We approximate the object $J(\pi)$ as another DNN, which is called a \textit{critic DNN}, whose parameter vector is denoted by $\theta_{Q}$. The parameter vector $\theta_{Q}$ is updated using the \textit{experience replay} and the \textit{target network technique} such as the \textit{deep Q-network algorithm} \cite{DQN}. These techniques can reduce correlation between experience data and make the learning performance stable, respectively.

The parameter vector $\theta_{Q}$ is updated by reducing the following \textit{critic loss function}.
\begin{eqnarray}
J(\theta_{Q})=E_{(z,a,z',r)\sim\mathcal{D}}\left[(Q_{\theta_{Q}}(z,a)-(r+\gamma V_{\theta_{Q}^{-}}(z')))^2\right].\label{critic_update}
\end{eqnarray}
The agent selects some experiences from a replay buffer $\mathcal{D}$ randomly for updates of $\theta_{Q}$. The value $V_{\theta_{Q}^{-}}(z')$ is computed by
\begin{eqnarray}
V_{\theta_{Q}^{-}}(z')=E_{a'\sim \pi_{\theta_{\pi}}}\left[Q_{\theta_{Q}^{-}}(z',a')-\alpha\log\pi_{\theta_{\pi}}(a'|z')\right],\nonumber
\end{eqnarray}
where $Q_{\theta_{Q}^{-}}$ is the \textit{target critic DNN}. The parameter vector $\theta_{Q}^{-}$ is slowly updated by the following \textit{soft update}.
\begin{eqnarray}
\theta_{Q}^{-}\leftarrow \xi\theta_{Q}+(1-\xi)\theta_{Q}^{-},\label{soft_update}
\end{eqnarray}
where $\xi>0$ is a sufficiently small positive constant. The parameter vector $\theta_{\pi}$ is updated by decreasing the following \textit{actor loss function}.
\begin{eqnarray}
J(\theta_{\pi})=E_{z\sim\mathcal{D},a\sim \pi_{\theta_{\pi}}}\left[\alpha\log(\pi_{\theta_{\pi}}(a|z))-Q_{\theta_{Q}}(z,a)\right].\label{actor_update}
\end{eqnarray}
The entropy temperature $\alpha$ is updated by decreasing the following loss function.
\begin{eqnarray}
J(\alpha)=E_{z\sim\mathcal{D}}\left[\alpha(-\log(\pi_{\theta_{\pi}}(a|z))-\mathcal{H}_{0})\right],\label{entropy_update}
\end{eqnarray}
where $\mathcal{H}_{0}$ is a lower bound. For example, in \cite{SAC}, $\mathcal{H}_{0}$ is set to $-\dim(\mathcal{U})$, where $\dim(\mathcal{U})$ denotes the dimension of the control action space $\mathcal{U}$. Actually, to keep $\alpha$ nonnegative after updates, we exponentiate the parameter.

\subsection{Preprocessing for extended states}
As $\tau$ becomes larger, the dimension of the extended state $z$ also increases. Thereafter, it is difficult for an agent to learn its policy because of the curse of dimensionality. Thus, we use a preprocess to decrease the dimension of the extended state \cite{Practical_STL_Q_learning}. Although the preprocess is proposed for grid world problems, it can also be applied to continuous control tasks. We introduce the flag value $f^{i}$ for each STL sub-formula $\phi_{i}$. 

\noindent \bf Definition 2: \rm For an extended state $z=[(x^{\tau})^{\top}\ (a^{d})^{\top}]^{\top}$, the flag value $f^{i}$ of an STL sub-formula $\phi_{i}$ is defined as follows:

\noindent (i) For $\phi_{i}=G_{[t_{s}^{i},t_{e}^{i}]}\varphi_{i}$,
\begin{eqnarray}
f^{i}=\max\left\{\frac{t_{e}^{i}-l+1}{t_{e}^{i}-t_{s}^{i}+1}\ \middle|\ l\in\{t_{s}^{i},...,t_{e}^{i}\}\hspace{0.5cm} \right.\nonumber\\ 
\biggl. \land(\forall l'\in\{l,...,t_{e}^{i}\},\ x^{\tau}[l']\models\varphi_{i}) \biggr\}.\label{G}
\end{eqnarray}

\noindent (ii) For $\phi_{i}=F_{[t_{s}^{i},t_{e}^{i}]}\varphi_{i}$,
\begin{eqnarray}
f^{i}=\max\left\{\frac{l-t_{s}^{i}+1}{t_{e}^{i}-t_{s}^{i}+1}\ \middle|\hspace{3cm} \right. \nonumber\\
\biggl.\hspace{2cm} l\in\{t_{s}^{i},...,t_{e}^{i}\}\land x^{\tau}[l]\models\varphi_{i}\biggr\}.
\label{F}
\end{eqnarray}

\begin{algorithm}               
\caption{Preprocessing of the extended state $z$}         
\label{preprocess}                          
\begin{algorithmic}[1]
\STATE \bf Input: \rm The extended state $z=[(x^{\tau})^{\top}\ (a^{d})^{\top}]^{\top}$ and the STL sub-formulae $\{\phi_{i}\}_{i=1}^{M}$.    
\FOR{$i=1,...,M$}   
\IF{$\phi_{i}=G_{[t_{s}^{i},t_{e}^{i}]}\varphi_{i}$}
\STATE Compute the flag value $f^{i}$ by Eq. (\ref{G}).
\ENDIF
\IF{$\phi_{i}=F_{[t_{s}^{i},t_{e}^{i}]}\varphi_{i}$}
\STATE Compute the flag value $f^{i}$ by Eq. (\ref{F}).
\ENDIF       
\ENDFOR
\STATE Set the flag state $\hat{f}=[\hat{f}^{1}\ \hat{f}^{2}\ ...\ \hat{f}^{M}]$ by Eq. (\ref{transformed}).
\STATE \bf Output: \rm The preprocessed state \\
$\hat{z}=[x^{\tau}[\tau-1]^{\top}\ \hat{f}^{\top}\ (a^{d})^{\top}]^{\top}$.    
\end{algorithmic}
\end{algorithm}

\begin{algorithm}
\caption{SAC-based algorithm to design a networked controller with network delays satisfying an STL specification}         
\label{alg1}                          
\begin{algorithmic}[1]                
\STATE Initialize the parameter vectors of main critic DNNs $\theta_{Q},\theta_{\pi}$. 
\STATE Initialize the parameter vector of a target critic DNN $\theta_{Q}^{-}$.
\STATE Initialize a replay buffer $\mathcal{D}$.
\FOR{$\text{Episode}=1,...,\text{MAX EPISODE}$}
\STATE Initialize the system state $x_{0}\sim p_{0}$.
\FOR{Discrete-time step $t=0,...,T$}
\IF{$t\ge d_{\text{sc}}$}
\STATE Receive the $(t-d_{\text{sc}})$-th observed state $x_{t-d_{\text{sc}}}$.
\STATE Construct the extended state $z_{t-d_{\text{sc}}}$.
\STATE Compute the next preprocessed state $\hat{z}_{t-d_{\text{sc}}}$ by \bf Algorithm 1\rm.
\IF{$t>d_{\text{sc}}$}
\STATE Compute the reward $r_{t-d_{\text{sc}}-1}=R_{z}(z_{t-d_{\text{sc}}-1})$.
\STATE Store the experience 
\begin{eqnarray}
(\hat{z}_{t-d_{\text{sc}}-1},a_{t-d_{\text{sc}}-1},\hat{z}_{t-d_{\text{sc}}},r_{t-d_{\text{sc}}-1})\nonumber
\end{eqnarray}
in the replay buffer $\mathcal{D}$.
\ENDIF
\STATE Determine the action $a_{t-d_{\text{sc}}}$ based on the state $\hat{z}_{t-d_{\text{sc}}}$.
\STATE Send the $(t-d_{\text{sc}})$-th control action $a_{t-d_{\text{sc}}}$ to the actuator.
\ENDIF
\STATE Sample $I$ experiences 
\begin{eqnarray}
\{(\hat{z}^{(i)},a^{(i)},\hat{z}'^{(i)},r^{(i)}) \}_{i=1,...,I}\nonumber
\end{eqnarray}
from the replay buffer $\mathcal{D}$ randomly.
\STATE Update the main DNNs $\theta_{Q},\theta_{\pi}$ by Eqs.\ (\ref{critic_update}) and (\ref{actor_update}). 
\STATE Update the target critic DNN $\theta_{Q}^{-}$ by Eq.\ (\ref{soft_update}).
\STATE Update the entropy temperature $\alpha$ by Eq.\ (\ref{entropy_update}).
\ENDFOR
\ENDFOR
\end{algorithmic}
\end{algorithm}
Note that $\max \emptyset =-\infty$ and the flag value represents the normalized time lying in $(0,1] \cup \{ -\infty \}$. Intuitively, for $\phi_i=G_{[t_{s}^{i}, t_{e}^{i}]} \varphi_i$, the flag value indicates the time duration in which $\varphi_i$ is always satisfied, whereas for $\phi_i=F_{[t_{s}^{i}, t_{e}^{i}]} \varphi_i$, the flag value indicates the instant when $\varphi_i$ is satisfied. The flag values $f^{i},\ i \in \{ 1,2 ,\dots, M\}$ calculated by Eqs. (\ref{G}) or (\ref{F}) are transformed into $\hat{f}^{i}$ as follows:
\begin{eqnarray}
\hat{f}^{i}= \left\{
\begin{array} {ll}
f^i -\frac{1}{2} & \mbox{if } f^i \not = -\infty, \\
-\frac{1}{2} & \mbox{otherwise}.
\end{array}\right.\label{transformed}
\end{eqnarray}
The transformed flag values $\hat{f}^{i}$ are used as inputs to DNNs to avoid positive biases of the flag  values and inputting $-\infty$ to DNNs. We compute the flag value for each STL sub-formula and construct a flag state $\hat{f}=[\hat{f}^{1}\ \hat{f}^{2}\ ...\ \hat{f}^{M}]^{\top}$, which is called \it preprocessing\rm. We use the preprocessed state $\hat{z}=[x^{\tau}[\tau-1]^{\top}\ \hat{f}^{\top}\ (a^{d})^{\top}]^{\top}$ instead of the extended state $z=[(x^{\tau})^{\top}\ (a^{d})^{\top}]^{\top}$. If $M\ll\tau$, we can decrease the dimension of the extended state as shown in Table\ \ref{table_dim_extended_state}. \bf Algorithm 1 \rm summarizes the above preprocessing.

\begin{table}[!t]
\renewcommand{\arraystretch}{1.3}
\caption{Dimension of extended state spaces}
\label{table_dim_extended_state}
\centering
\begin{tabular}{c|c|c}
\hline
\bfseries  & Without Preprocessing $z$ & With Preprocessing $\hat{z}$\\
\hline
Dimension & $\tau n_x+d n_u$ & $n_x + M + d n_u$\\
\hline
\end{tabular}
\end{table}

\subsection{Algorithm}
We propose an SAC-based algorithm presented in \bf Algorithm 2\rm. From lines 1 to 3, we initialize the parameter vectors of DNNs and a replay buffer $\mathcal{D}$. In line 5, we initialize the state of the system. From lines 6 to 22, the agent interacts with the system and learns its policy for an episode. In line 8, at $t\ (\ge d_{\text{sc}})$, the agent receives the state $x_{t-d_{\text{sc}}}$. In line 9, the extended state $z_{t-d_{\text{sc}}}$ is constructed using $x_{t-d_{\text{sc}}}$, $z_{t-d_{\text{sc}}-1}$, and $a_{t-d_{\text{sc}}-1}$. In line 10, the preprocessed state $\hat{z}_{t-d_{\text{sc}}}$ is computed by \textbf{Algorithm 1}. In line 12, if $t>d_{\text{sc}}$, the reward $r_{t-d_{\text{sc}}-1}$ is computed by Eq.\ (\ref{reward_function}). In line 13, the agent stores the experience $(\hat{z}_{t-d_{\text{sc}}-1},a_{t-d_{\text{sc}}-1},\hat{z}_{t-d_{\text{sc}}},r_{t-d_{\text{sc}}-1})$ to the replay buffer $\mathcal{D}$. In line 15, the agent determines an exploration action $a_{t-d_{\text{sc}}}$ based on the preprocessed state $\hat{z}_{t-d_{\text{sc}}}$. In line 16, the agent sends $a_{t-d_{\text{sc}}}$ to the actuator. From lines 18 to 21, the agent updates the DNNs. In line 18, the agent samples $I$ past experiences $\{(\hat{z}^{(i)},a^{(i)},\hat{z}'^{(i)},r^{(i)})\}_{i=1}^{I}$ from the replay buffer $\mathcal{D}$ randomly. From lines 19 to 21, the agent updates the parameter vectors of the DNNs based on the SAC algorithm. 

\section{Example}
\begin{figure}[h]
\begin{center}
  \includegraphics[width=8.5cm]{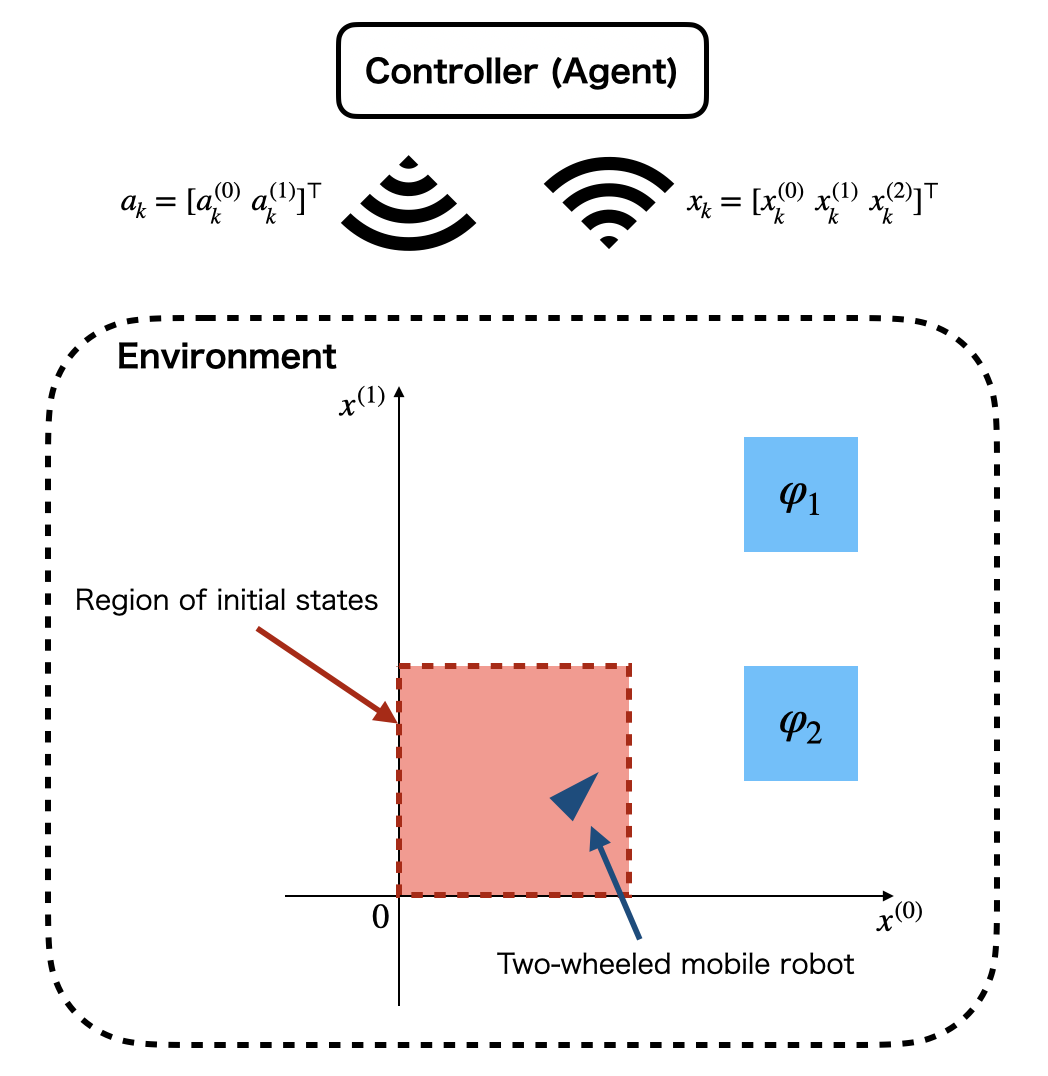}
  \caption{Environment of the example. The agent learns the optimal policy for the two-wheeled mobile robot under the STL constraint.}
  \label{Example_pic}
\end{center}
\end{figure}

Consider a two-wheeled mobile robot in the environment shown in Fig.\ \ref{Example_pic}. A discrete-time model of the robot is described by
\begin{eqnarray}
\begin{bmatrix}
	x_{t+1}^{(0)}\\
	x_{t+1}^{(1)}\\
	x_{t+1}^{(2)}	
\end{bmatrix}=\begin{bmatrix}
	x_{t}^{(0)} + \Delta u_{t}^{(0)}\cos(x_{t}^{(2)})\\
	x_{t}^{(1)} + \Delta u_{t}^{(0)}\sin(x_{t}^{(2)})\\
	x_{t}^{(2)} + \Delta u_{t}^{(1)}
\end{bmatrix}+\Delta_{w}\begin{bmatrix}
	w_{t}^{(0)}\\
	w_{t}^{(1)}\\
	w_{t}^{(2)}
\end{bmatrix},\label{example}
\end{eqnarray}
where $x_{t}=[x_{t}^{(0)}\ x_{t}^{(1)}\ x_{t}^{(2)}]^{\top}\in\mathbb{R}^{3}$, $u_{t}=[u_{t}^{(0)}\ u_{t}^{(1)}]^{\top}\in\mathbb{R}^{2}$, and $w_{t}=[w_{t}^{(0)}\ w_{t}^{(1)}\ w_{t}^{(2)}]^{\top}\in\mathbb{R}^{3}$. $w_{t}^{(i)},\ i\in\{0,1,2\}$ is sampled from a standard normal distribution $\mathcal{N}(0,1)$. In the simulation, we assume that $\Delta=0.1$ and $\Delta_{w}=0.01I$, where $I$ is the identity matrix. The initial state of the system is sampled randomly in the region $0\le x^{(0)}\le 2.5,\ 0\le x^{(1)}\le 2.5,\ -\pi/2\le x^{(2)}\le \pi/2$. 

In this example, we consider the following STL formula.
\begin{eqnarray}
\Phi &=& G_{[0,900]}(F_{[0,99]}\varphi_{1}\land F_{[0,99]}\varphi_{2}),\label{specification_1}
\end{eqnarray}
where
\begin{eqnarray}
\varphi_{1} &=& ((3.75\le x^{(0)}\le 5)\land(3.75\le x^{(1)}\le 5)),\nonumber\\
\varphi_{2} &=& ((3.75\le x^{(0)}\le 5)\land(1.25\le x^{(1)}\le 2.5)),\nonumber
\end{eqnarray}
that is, the length of $x^{\tau}$ is $\tau=100$. It is assumed that $d_{\text{sc}}=3$ and $d_{\text{ca}}=4$, where these values are unknown, but we know that $d_{\text{sc}}\le 5$ and $d_{\text{ca}}\le 5$ beforehand. Then, the length of the past control action sequence $a^{d}$ is $d=10$. In all simulations, the DNNs have two hidden layers, all of which have 256 units, and all layers are fully connected. To mitigate the positive bias in the update of $\theta_{\pi}$, the clipped double Q-learning technique \cite{TD3} is adopted for $Q_{\theta_{Q}}(\hat{z},a)$. The activation functions for the hidden layers and the outputs of the actor DNN are the rectified linear unit functions and hyperbolic tangent functions, respectively. We normalize $x^{(0)}$ and $x^{(1)}$ as $x^{(0)}-2.5$ and $x^{(1)}-2.5$, respectively. The size of the replay buffer $\mathcal{D}$ is $1.0\times10^{5}$, and the size of the mini-batch is $I=64$. We use \textit{Adam} \cite{Adam} as the optimizers for all main DNNs and the entropy temperature. The learning rates of all optimizers multiplier are $3.0\times10^{-4}$. The soft update rate of the target network is $\xi=0.01$. The discount factor is $\gamma=0.99$. The target for updating the entropy temperature $\mathcal{H}_{0}$ is $-2$. The STL-reward parameter is $\beta=100$. The agent learns its control policy for $6.0\times10^{5}$ steps. The initial entropy temperature is $1.0$. For performance evaluation, we introduce the following two indices:
\begin{itemize}
\item  a \textbf{learning curve} shows the mean of returns $\sum_{k=0}^{T-d_{\text{sc}}-d_{\text{ca}}}\gamma^{k}R_{z}(z_{k})$ for 100 trajectories, and
\item a \textbf{success rate} shows the number of trajectories satisfying the given STL constraint for 100 trajectories.
\end{itemize}
We prepare $100$ initial states sampled from $p_0$ and generate $100$ trajectories using the learned policy for each evaluation. All simulations are run on a computer with AMD Ryzen 9 3950X 16-core processor and 32GB of memory and are conducted using the Python software.

\subsection{Effect of network delays}
\begin{figure}[h]
\begin{center}
  \includegraphics[width=8.5cm]{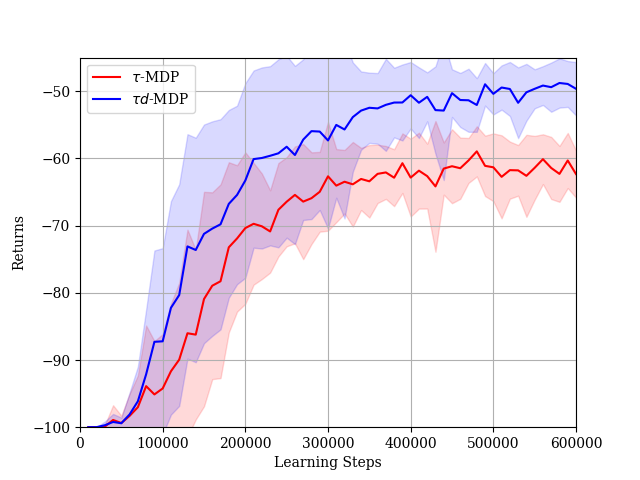}
  \caption{Learning curves for the $\tau$-MDP case and $\tau d$-MDP case. The solid curve and the shade represent the average results and standard deviations over 15 trials with different random seeds, respectively.}
  \label{Example_fig_1}
\end{center}
\end{figure}

\begin{figure}[h]
\begin{center}
  \includegraphics[width=8.5cm]{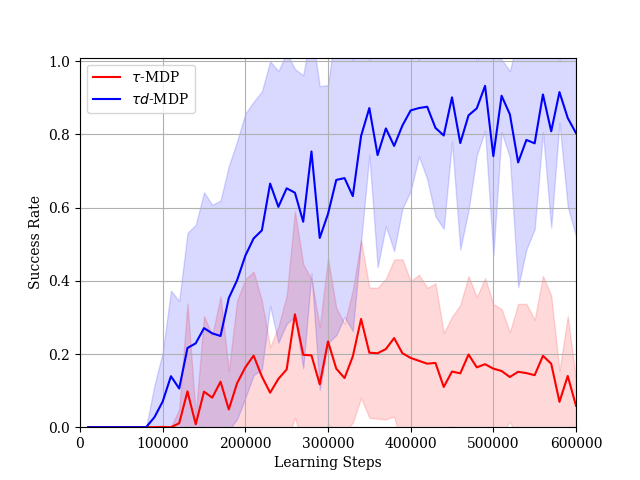}
  \caption{Success rates for the $\tau$-MDP case and the $\tau d$-MDP case. The solid curve and the shade represent the average results and the standard deviations over 15 trials with different random seeds, respectively.}
  \label{Example_fig_2}
\end{center}
\end{figure}

In this section, we demonstrate the effect of using past control actions as a part of an extended state, where we use the preprocessing introduced in Section IV.\ B. The learning curves and the success rates for the $\tau$-MDP case (without past determined actions) and the $\tau d$-MDP case (with past determined actions) are shown in Figs.\ \ref{Example_fig_1} and \ref{Example_fig_2}, respectively. If we do not use past determined actions, the obtained returns become high as the agent updates its policy, as shown in Fig.\ \ref{Example_fig_1}, but the success rate of the learned policy is not increasing as shown in Fig.\ \ref{Example_fig_2}. Conversely, if we use past determined actions, the agent can learn the policy that has a high success rate for the given STL specification. This result concludes that the agent needs not only past states but also past determined actions to learn the policy satisfying the given STL specification with network delays.

\subsection{Effect of preprocess}
\begin{figure}[h]
\begin{center}
  \includegraphics[width=8.5cm]{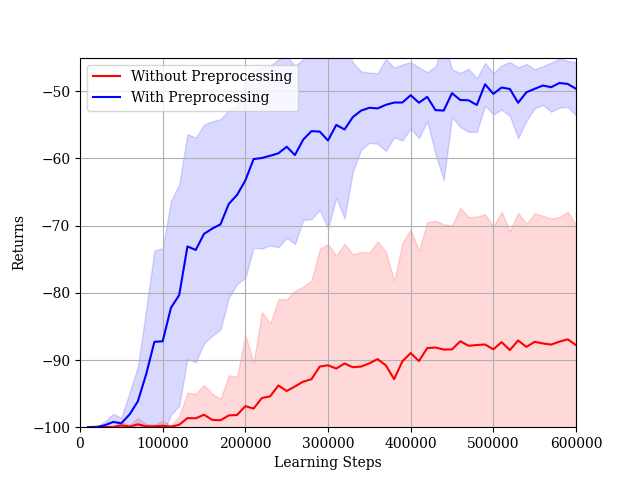}
  \caption{Learning curves for the cases with and without preprocessing. The solid curve and the shade represent the average results and the standard deviations over 15 trials with different random seeds, respectively.}
  \label{Example_fig_3}
\end{center}
\end{figure}

\begin{figure}[h]
\begin{center}
  \includegraphics[width=8.5cm]{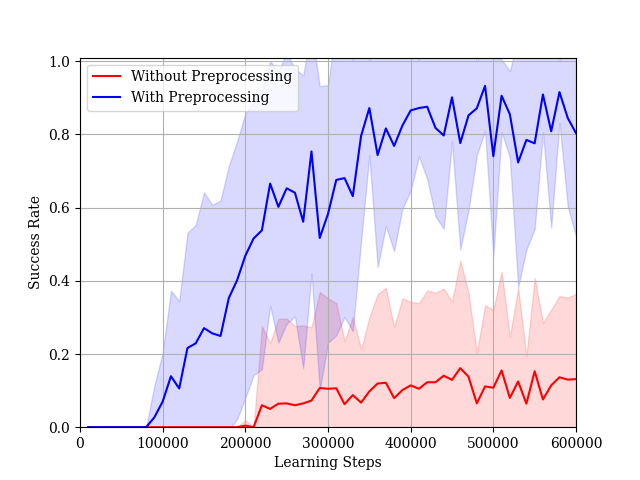}
  \caption{Success rates for the cases with and without preprocessing. The solid curve and the shade represent the average results and the standard deviations over 15 trials with different random seeds, respectively.}
  \label{Example_fig_4}
\end{center}
\end{figure}

In this section, we show the improvement in the learning performance by preprocessing. In the case without preprocessing, the dimension of the extended state is $320$ and, in the case with preprocessing, the dimension of the extended state is $25$. As shown in Fig.\ \ref{Example_fig_3}, the agent cannot improve the performance of its policy without preprocessing. Then, the learned policy has  a low success rate as shown in Fig.\ \ref{Example_fig_4}. Conversely, in the case with preprocessing, the agent can learn a policy that obtains high returns and a high success rate. The result concludes that preprocessing is necessary for our proposed method under the STL specification with a large $\tau$ to mitigate curse of dimensionality.

\section{Conclusion}
We proposed a DRL-based networked controller design for a given STL specification with network delays. Subsequently, we introduced an extended MDP, which is called a $\tau d$-MDP, and proposed a DRL algorithm to design the networked controller. Through numerical simulations, we demonstrated the performance of the proposed method. On the other hand, for some STL specifications, the reward may be sparse. Additionally, the syntax in this study is the restrictive compared with the general STL syntax \cite{STL}. Solving these issues is an interesting direction for future work. As a practical problem, an extension of the proposed method to a system with network delays that fluctuate randomly is also a future work. 

\section*{Acknowledgment}
This work was partially supported by JST-ERATO HASUO Project Grant Number JPMJER1603, Japan and JSPS KAKENHI Grant Number JP21J10780, Japan

\end{document}